\documentstyle[aps,manuscript]{revtex}
\begin{document}
\baselineskip=23pt 
\draft
\title {Self-Organized Segregation within an Evolving Population} 
\author{N.F. Johnson$^1$, P.M. Hui$^2$, R. Jonson$^1$ and T.S. Lo$^2$}  
\address {$^1$ Department of Physics, University of Oxford,\\  Parks Road,
Oxford, OX1 3PU, England, United Kingdom}
\address {$^2$ Department of Physics, The Chinese University of Hong Kong,\\
Shatin, New Territories, Hong Kong}  
%
\date{\today}
\maketitle


\begin{abstract}

An evolving population, in which individual members (`agents') adapt their
behaviour according to past experience, is of central importance to many
disciplines.  Because of their limited knowledge and capabilities, agents
are
forced to make decisions based on inductive, rather than deductive,
thinking. 
We show that a population of competing agents with similar capabilities and
knowledge will tend to  self-segregate  into opposing groups characterized
by
extreme behavior.  Cautious agents perform poorly  and tend to become
rare.
\vskip\baselineskip
\vskip\baselineskip

\noindent {\bf To appear in Phys. Rev. Lett. (1999)}
\vskip\baselineskip
\vskip\baselineskip

\end{abstract}


\newpage

In physical systems, simple rules applied to a set of $N\geq 3$ interacting
objects can give rise to complex dynamical behavior. Although generally
intractable analytically, such problems are  simplified considerably by the
fact
that the inter-particle interactions are typically instantaneous,
time-independent and decrease with increasing particle-particle separation. 
An
arguably more complex problem which is of central importance in social,
economic
and biological sciences
\cite{bak,Arthur,physics,epstein,Us}, is that of an evolving
population in which individual members (`agents') adapt their interactions,
and
hence behavior, according to their past experiences. Even the two-player
($N=2$)
prisoners' dilemma game played by memory-less agents on a lattice, has been
shown numerically to yield rich spatio-temporal patterns
\cite{Nowak}. Evolutionary game theory has been applied to such many-agent
systems
\cite{Sigmund}. However it is well-known that such analysis provides
little insight into the system's dynamics.

Of particular interest is the situation where  agents repeatedly compete for
a
limited resource, or to be in a minority. Rush-hour drivers, facing the
nightly
choice between two alternative routes home, wish to choose the route
containing
the minority of traffic
\cite{Huberman}. In financial markets, more buyers than sellers implies
higher
prices; hence it is better for a trader to be in the minority group of
sellers.
Animals (salesmen) foraging for food (customers) will do better if they hunt
in
areas with fewer competitors. Regular attendees at a popular bar may try to
avoid over-crowded evenings \cite{Arthur,Us}.  More generally, the problem
of
how to flourish in a population of equally ambitious people with similar
capabilities, but where there are typically more losers than winners, is one
that many people face daily.  

Here we introduce a simple, yet realistic, model for such an evolving
population
containing adaptive agents who compete to be in the minority.   Only {\em
partial} information about the system is available to the agents and no a
priori
`best' strategy exists: agents are hence forced to make decisions based on
inductive, rather than deductive, thinking. Each agent tries to learn from
its
past mistakes and will adjust its strategy in order to survive.  We find
that a
population of such agents with similar capabilities will tend to polarize 
itself
into opposing groups. Although a large number of possible strategies exist,
the
most successful agents are those who behave in an extreme way.

Inspired by Ref. \cite{Zhang}  we consider the model of an odd number $N$ of
agents  repeatedly choosing whether to be in room `0' or room `1'. These
agents
could be daily traders or rush-hour drivers: choosing room `0' denotes
choosing
to buy a given asset or choosing to take route A, respectively, while  `1'
denotes choosing to sell the asset or choosing to take route B. After every
agent has independently chosen a room, the winners are those in the minority
room, i.e. the room with fewer agents. The `output' for each time-step  is a
single binary digit denoting the winning room. Each agent is given a
bit-string
of length $m$ containing the previous
$m$ outcomes. Each agent also has access to a common
register or `memory' containing the outcomes from the most recent 
occurrences
of all
$2^m$ possible bit-strings of length $m$. Consider
$m=3$; denoting (xyz)w as the
$m=3$ bit-string (xyz) and outcome w, an example memory would comprise
(000)1,
(001)0, (010)0, (011)1, (100)0, (101)1, (110)0, (111)1. Following a run of
three
wins for room 0 in the recent past, the winning room was subsequently 1.  
Faced
with a given bit-string of length $m$, it might seem sensible for an agent
to
simply predict the same outcome as that registered in the memory. The agent
will
hence choose room 1 following the next 000 sequence. If 0 turns out to be
the
winning room, the entry (000)1 in the memory is replaced by (000)0. If all
$N$ agents act in this way, however, the system will be inefficient since
all
agents will choose the same room and will hence lose; all the agents are
spotting the same trends and assuming that they will continue indefinitely.
Because of this, the trend fails to continue. The critical quality of a
successful financial trader, for example, is the ability to follow a trend
as
long as it is valid, but to correctly predict when it will end. Hence we
assign
each agent a single number or `strategy' $p$: following a given $m$-bit
sequence,
$p$ is the probability that the agent will choose the same outcome as that
stored in the memory, i.e. he will follow the current prediction, while
$1-p$ is the probability he will choose the opposite, i.e. he will reject
the
current prediction. Using the example memory, the agent (e.g. trader or
driver)
will choose 1 (e.g. sell or take route B) with probability
$p$ after spotting the sequence 000, and $0$ (e.g. buy or take route A) with
probability
$1-p$.

Each time an agent gets into the minority (majority) room, he gains (loses)
one
point. If the agent's score  falls below a value $d<0$, then his strategy is
modified, i.e. the agent gets a new $p$ value which is chosen with an equal
probability from a range of values, centered on the old
$p$, with a width equal to $R$. Hence $d$ is the number of times (or the
amount
of money) a driver (or trader) is willing to be wrong (or to lose) before
modifying his/her strategy. Although this is a fairly crude
`learning' rule as far as machines are concerned \cite{Sutton}, in
our experience it is not too dissimilar from the way that humans actually
behave
in practice.  Since
$0\leq p
\leq 1$, we can for simplicity enforce reflective boundary conditions. Our
conclusions do not depend on the particular choice of boundary conditions
(see
Fig. 1). Upon strategy modification, the agent's score is reset to zero.
Changing
$R$ allows the way in which the agents learn to be varied. For $R=0$, the
strategies will never change (though the memory will).  If $R=2$, the
strategies
before and after modification are uncorrelated. For small $R$, the new
$p$ value is close to the old one. 

As agents (e.g. traders or drivers) are constantly attempting to do the
opposite
of other agents, a reasonable expectation is that they should eventually
organise
themselves so that their strategies are evenly spread within
$0\leq p \leq 1$. Alternatively, given that no a priori best strategy
exists,
one might expect that agents would be ambivalent as to whether a present
trend
will continue, and hence cluster around
$p=\frac{1}{2}$. Surprisingly, the opposite is true.  Figure 1(a) shows the
frequency distribution $P(p)$ at various times.
The distribution $P(p)$
eventually becomes peaked around
$p=0$ and
$1$ (solid line) regardless of the initial $P(p)$ distribution; these $p$
values
respectively correspond  to always or never following what happened last
time.
The lifespan
$L(p)$, defined as the average length of time a strategy
$p$ survives between modifications, shows similar behavior
(solid line in Fig. 1(b)). Henceforth we denote $P(p)$ and $L(p)$ as 
representing the long-time limits (solid lines).
If we consider the game simply as a random walk, with individual agents
deciding
randomly which room to choose, we would expect the mean number in room 0 or
1 to
be $N/2$ with a standard deviation of
$\sqrt{N/4}$.  At each timestep, the net number of points awarded will
therefore
be $-\sqrt{N}$. The average lifespan would  be $d\sqrt{N}$.  The observed
average lifespan is indeed   proportional to $d\sqrt{N}$.   However the
average
value of the $L(p)$ in Fig. 1(b) (solid line) is larger than
$d\sqrt{N}$ by a factor of approximately 2 for
$d=-4$, confirming that the agents are organizing themselves better than
randomly.  Furthermore, the root-mean-square (rms) separation of the
strategies
is higher than the value for uniform $P(p)$, indicating the desire of agents
to
do the opposite of the majority. It increases with $N$ due to increased
possibilities for self-organization. Even when $R$ is large,  and the
strategy
values are hence picked randomly upon modification, the rms strategy
separation
remains high. The rms strategy separation and the average value of 
$L(p)$ are typically maximal at $R\sim 0.5$; this is a
compromise between a lack of learning when 
$R\sim 0$ and excessive strategy modification for large $R$.  
We also  
note that the standard deviation of the actual attendance time-series 
for room 0 (or room 1) is less than that obtained for agents choosing 
via independent coin-tosses:  this again confirms that the system is 
organizing itself better than random. 

Varying the length of the bit-string $m$ has little effect on $P(p)$ and
$L(p)$: since all agents have similar capabilities and available
information,
these benefits tend to cancel out.  It is what each agent decides to do with
the
common knowledge which matters ($p=0,1$ agents outperform
$p=\frac{1}{2}$ agents).  Similarly 
if the memory is not updated dynamically according
to the recent outcomes as discussed earlier, but is instead kept
constant (i.e. time-independent) or is randomly chosen at each time-step,
then
$P(p)$ and
$L(p)$ are also essentially unchanged. Once again, the
memory is common to all agents and hence all agents agree on the current
prediction: no agent hence has any relative advantage in terms of available
information
\cite{Cavagna}.
It has been shown for the basic minority game \cite{Savit2}, in contrast 
to the claim in Ref. \cite{Cavagna}, that the memory is relevant 
since it can introduce hidden correlations into the winning-room 
time series.  This point will be discussed in detail for the present 
model elsewhere. 

We now provide some analytic analysis. The simplest example of our system
contains $N=3$ agents $i,j,k$ with brain-size $m$ and three discrete $p$
values
$p=0,\frac{1}{2},1$. (The fact that $N<3$ is impossible suggests that our
system
contains the level of complexity typically associated with 3-body, versus
2-body, problems).  All agents agree on the current prediction (say 0).
Agent
$i$ will choose
$0$ or $1$ with probability $p_i$ and
$1-p_i$ respectively.  Likewise for $j$ ($p_j$) and $k$ ($p_k$). The
$2^3$ possible decisions for $ijk$ are 000, 001, 010, 100, 110, 101, 011,
111. 
There are
$3^3=27$ possible configurations $(p_i,p_j,p_k)$. For a given
$(p_i,p_j,p_k)$, the 8 possible decisions yield the expected gain for the
agents. For example for $(p_i,p_j,p_k)=(0,0,\frac{1}{2})$, $i$ and $j$ both
choose 1 while $k$ chooses 0 with probability $\frac{1}{2}$. Hence $k$ wins
with
probability $\frac{1}{2}$ whereas $i$ and $j$ both lose. The net number of
points gained per agent per turn, given by the points awarded minus the
points
deducted, is $-1$ for $i$, $-1$ for $j$, and $0$ for $k$. The total is hence
$-2$. Given that the  maximum is $-1$ (there is a maximum of one winner) we
see
that $(0,0,\frac{1}{2})$ is not optimal. 

Table 1 shows the various configuration types, or classes. The last column
shows
the average points per agent: 
$[-\frac{1}{2}]$ for class i) implies the average agent loses $-\frac{1}{2}$
point per turn, and would hence modify its strategy after time
$2d$.  Such strategy modification allows the system to sample the 27
configurations. Classes vi), vii) and viii) are optimal, having maximum
points.
To obtain the average distribution
$P(p)$ and $L(p)$, we must average over all 27 configurations. Since some
classes are more favourable (i.e., more points) we should weight the
distributions in an appropriate way. In the extreme case of large weighting,
we
include only the optimal classes vi), vii) and viii), yielding
$P(0):P(\frac{1}{2}):P(1)$$=2.5:1:2.5$ and
$L(0):L(\frac{1}{2}):L(1)$$=5:1:5$.  For zero weighting, we instead consider
the
system as visiting all configurations with equal probability regardless of
points gained per agent; such a zero-weight averaging is similar to that for
the
microstates in a gas within the microcanonical ensemble and yields
$P(0):P(\frac{1}{2}):P(1)$$=1:1:1$ and $L(0):L(\frac{1}{2}):L(1)$$=1:1:1$. 
For
an intermediate case, whereby all classes are weighted by the average points
per
agent, we obtain
$P(0):P(\frac{1}{2}):P(1)$$=1.1:1:1.1$ and
$L(0):L(\frac{1}{2}):L(1)$$=1.5:1:1.5$. In fact, any sensible weighting
which
favours the more profitable configurations yields a non-uniform $P(p)$ and
$L(p)$ as observed numerically. This implies that the population, 
by self-segregating, has also managed to 
self-organize itself around the most profitable configurations. 
We emphasize that the system is
dynamic since the membership of the various configurations is constantly
changing ($i$,
$j$ and $k$ inter-diffuse) but $P(p)$ remains essentially constant. For
general
$N$ we can loosely think of $i,j,k$ as three equal-size groups of
like-minded
agents.   

In summary, we have shown that an evolving population of agents with similar
capabilities and information will self-segregate. To flourish in such a
population, an agent should behave in an extreme way ($p=0$ or $p=1$).

We thank D. Challet and P. Binder for discussions.

\vskip\baselineskip

\newpage

\begin{table}

\label{table1}
\caption{Configuration classes showing the distribution of the three agents
(each denoted by x) and the average points awarded per time-step for each
strategy-value $p$. Also given are the number of distinct configurations per
class, and the average number of points per agent per time-step.}
\vskip\baselineskip
\vskip\baselineskip
\vskip\baselineskip

\begin{tabular}{ccccccc} Class & $p=0$ & $p=1/2$ & $p=1$ & No. configs. &
Ave
pts/agent \\
\tableline i) & \_ & xxx[-1/2][-1/2][-1/2] & \_ & 1 & [-1/2] \\ ii) &
x[-1/2] &
xx[-1/2][-1/2] & \_ & 3 & [-1/2] \\ iii) & xx[-1][-1] & x[0] & \_ & 3 &
[-2/3] \\
iv) & xxx[-1][-1][-1] & \_ & \_ & 1 & [-1] \\ v) & \_ & \_ & xxx[-1][-1][-1]
& 1
& [-1] \\ vi) & x[1] & \_ & xx[-1][-1] & 3 & [-1/3] \\ vii) & xx[-1][-1] &
\_ &
x[1] & 3 & [-1/3] \\ viii)& x[0] & x[-1] & x[0]& 6 & [-1/3] \\ ix)& \_ &
xx[-1/2][-1/2] & x[-1/2] & 3 & [-1/2] \\ x) & \_ & x[0] & xx[-1][-1] & 3 &
[-2/3] \\
\end{tabular}

\end{table}

\newpage

\centerline{\bf Figure Captions}

\bigskip

\noindent Figure 1:  Distribution of (a) strategies $P(p)$. At $t=0$, 
$P(p)$ was chosen to be flat. Dashed line shows
$P(p)$ at intermediate times. Solid line shows $P(p)$ at large
times. (b) Corresponding lifespans
$L(p)$. The parameters
$R=0.2$,
$N=101$,
$d=-4$ and
$m=3$. Dotted lines show the long-time distributions using periodic  (as
opposed to reflective) boundary conditions.

\bigskip

\end{document}